

\font\titlefont=cmbx12 at 20pt


\footline{\ifnum\pageno>1\hss\lower.25in\hbox{\tenrm\folio}\hss\fi}


\def\topinfo#1;#2;{\rightline{#1}\par\rightline{#2}}

\def\abstract{\centerline{\bf Abstract}\medskip
\leftskip 40pt\rightskip 40pt\noindent}

\def\title#1{\centerline{\bf #1}}

\def\section#1{\noindent{\bf #1}\leftskip 0pt\rightskip
0pt\bigskip}

\def\subsection#1{\noindent{\bf #1}\medskip}


\def\grm{m_{3/2}}
\def\gam{m_{1/2}}
\def\mtop{m_{\rm t}}
\def\lamtop{\lambda_{\rm t}}
\def\mpl{\rm M_{pl}}
\def\mw{\rm M_w}

\def\mx{\rm M_X}
\def\gev{\rm GeV}
\def\tev{\rm TeV}


\def\vev#1{\langle #1 \rangle}
\def\order#1{{\cal O}(#1)}

\def\mod#1{\vert #1 \vert}


\def\put #1 #2 at (#3,#4) {\offinterlineskip
  \vskip-#4\noindent\rlap{\hskip#3
  \hbox to 0pt{\if#1r\else\hss\fi\smash{#2}\if#1l\else\hss\fi}}%
  \vskip#4}

\def\bpict #1 (#2 by #3) labels #4 left #5 right #6 up #7 down #8 sc #9 epict
  {\setbox0\hbox{\hskip #5\vtop to #3{\hsize=#2
  \hrule width #2 height 0pt depth 0pt
  \vfill \special{picture #1 scaled #9}#4 }\hskip #6}%
  \dimen20=\ht0\advance\dimen20 by #7\ht0=\dimen20
  \dimen20=\dp0\advance\dimen20 by #8\dp0=\dimen20
  \box0}


\topinfo MIU-THP 95/73;October 1995;
\vskip 1.5in

\title{\titlefont Dynamical Determination of the}
\bigskip
\title{\titlefont Fundamental Couplings}
\vskip 1.2in

\centerline{{\bf J.\ Hagelin, S.\ Kelley}\footnote{$^1$}{email
stephen\_kelley@telegroup.com} {\bf and Sunil Rawal}\footnote{$^2$}{email
srawal@miu.edu}} \medskip
\centerline{\it Department of Physics}
\centerline{\it Maharishi University of Management}
\centerline{\it Fairfield, IA 52557}
\vskip 1.2in

\abstract
\noindent
We demonstrate that supergravity models containing the Standard Model,
dilaton and modulus naturally lead to dynamical symmetry breaking with
excellent phenomenology. We assume primordial supersymmetry
breaking in the form of a constant contribution to the
superpotential. String inspired relations link fundamental couplings to the
dilaton vev. We specialize to a class of models inspired by the
$4$-$D$ fermionic string. Non-renormalizable terms in the
superpotential naturally produce the Higgs
mixing parameter $\mu$ suitable for our mechanism. We discuss extensions and
limitations of our approach.

\vskip0.8in\noindent
 Submitted to {\it Physical Review Letters}

\vfil\eject

\leftskip 0pt\rightskip 0pt

A fairly complete picture of physics from
the Planck scale to the weak scale can be painted in the framework
of string theory. However, although possible mechanisms for
supersymmetry breaking and fixing the vevs of moduli fields exist,
these issues remain unclear.  Perhaps the most complete and popular
mechanism relies on gaugino condensation in the hidden sector [1].
This approach has at least two major phenomenological
problems: first, the vacuum energy without fine-tuning is
$\order{\Lambda^4}$ where $\Lambda$ is the condensation scale and second, a
large
ratio for $m_0/\gam$ is uncomfortable in the light of experimental bounds on
the
gaugino mass [2] and the hierarchy problem [3].

In contrast, no-scale models have been successful
in breaking a flat modulus direction by radiative corrections while
giving a vacuum energy of $\grm^4$ and small ratios for $m_0/\gam$
[4]. However a no-scale mechanism for fixing the dilaton vev has
never been demonstrated. The dilaton is of particular interest since
it determines the unified gauge coupling [5]
$$g^{-2}(M_X)=\vev{S}.\eqno(1)$$
In this paper, we present a mechanism for dynamically determining the
vev of the dilaton field via radiative symmetry breaking.  Section 1
demonstrates this mechanism using a toy supergravity model.
Section 2 considers the mechanism in
the light of a specific class of string models inspired by $4D$
free-fermionic string constructions.  Section 3 outlines some
additional considerations and questions.
\bigskip
\section{1. A Simple Supergravity Model}
The simplest supergravity model which includes the dilaton, $S$, and
modulus field, $T$, and is capable of reproducing the Supersymmetric Standard
Model (SSM) with vacuum energy $\order{\grm^4}$ is constructed from the
K\"ahler function
$$G=-\ln{(S+\bar{S})}-2\ln{(T+\bar{T})}+z_i\bar{z}^i+\ln\mod{W}^2
\eqno(2)$$ where the $z_i$ are the chiral fields of the SSM (a bar
denotes complex conjugate). Generic string considerations [5] indicate
that the gauge kinetic function at tree-level is given by
$$f_{ab}=\delta_{ab}S.\eqno(3)$$
In addition, we take the superpotential
$$W=w+\mu H_1 H_2+\lamtop H_2 Q_3 U_3^c.\eqno(4)$$
where we have included a constant term $w$ coming from tree-level
supersymmetry breaking in the string (for instance from a generalized
compactification scheme [6]) and a Higgs mixing parameter $\mu$ whose
origin is considered to be unknown at this point and will be
considered later.

The model leads to a one-loop
corrected Planck-scale scalar potential in Landau gauge [7]
$$V={1\over 8}{(g_y^2 +
g_2^2)(\mod{H_1}^2 + \mod{H_2}^2)^2 + m_1^2 \mod{H_1}^2 + m_2^2
\mod{H_2}^2 - (B\mu H_1H_2} + c.c) + {1\over{64\pi^2}} {\rm Str}
{\cal M}^4 \bigl(\ln{{\cal M}^2\over Q^2} - {3\over 2}\bigr) + \eta
\grm^4\eqno(5)$$
where we regard $\eta$ and $\mu$ at the unification scale $\mx$ as
free parameters. The soft supersymmetry breaking
parameters coming from the model are simply calculated as
$$A(\mx)=3\grm,\quad B(\mx)=\Bigl(2-(S+\bar S){\partial\ln\mu\over\partial
S}\Bigr)\grm,\quad m_0=\gam=\grm\eqno(6)$$
and the gravitino mass $\grm$ is
$$\grm^2=e^{\vev{G}}={\vev{\mod{W}^2}\over
\vev{(S+\bar{S})(T+\bar{T})^2}}.\eqno(7)$$

The construction of the
low-energy scalar potential follows from the standard
one-loop RGEs. For the Higgs mass terms [8], we include only the
contributions of gauge couplings and the top-Yukawa for simplicity.
For the cosmological term [7] and supertrace terms
[9] we use analytic forms dependent upon the gauge coupling contributions of
squarks and gluinos. We have taken no thresholds in running the equations to
$M_z$.

In treating $\mu$ as of unknown origin, we can consider it to
depend only on the dilaton field, and further that its rescaled
form is given by
$$\hat\mu=g\mu(S).\eqno(8)$$
In this case, we can take its first and second
derivatives with respect to the dilaton field as two more free
parameters. We take the rescaled form of the
top-Yukawa at the unification scale as
$$\lamtop(\mx)=g^2\eqno(9)$$
and hence has no $T$ dependence. Motivation for these
choices will be given in Section 2.

The occurrence of the gravitino mass in the parameters $(6)$ implies
that the minimization with respect to the Higgs fields results in
$$\vev{H_{1,2}}=\tilde H_{1,2}(\tilde\mu)\grm\eqno(10)$$
where $\tilde H$ is a dimensionless function of
$\tilde\mu\ (=\mu/\grm)$ only. We can determine $\tan\beta$ and $\grm$
as functions of $\mu$ by using the known value of $\mw$:
$$\mw^2=g_2^2 \vev{H_1}^2(1 +
\tan^2\beta)\quad{\rm with}\quad \tan\beta={\vev{H_2}\over
\vev{H_1}}.\eqno(11)$$

Since the only appearance of the modulus field $T$ is in $\grm$,
minimizing in this field gives the ``no-scale condition'' of [10]:
$${\partial V\over\partial\grm}=0\eqno(12)$$
which determines a relation between $\eta(\mx)$, $\mu$ and
$\vev{S}$. This result implies that the minimization in $S$ only
involves explicit $S$ dependence coming from dilaton dependent
boundary conditions at the unification scale:
$$5/3g_1^2=g_2^2=g_3^2=\lamtop={1\over{\rm Re}
S}\quad{\rm at\ }\mx.\eqno(13)$$
A choice of $\partial\hat\mu/\partial S$ determines a value for $B$ through
(6).
The condition $\partial V/\partial S=0$ at a value of the dilaton field which
reproduces $\alpha_X=1/24$ then determines $\hat\mu$. Stability of the theory
must
then be verified for that choice of $\partial\hat\mu/\partial S$ since the
Hessian
involves non-trivial mixing of $S$ and $H$ derivatives. Table 1 indicates a
range
of $B$ for which consistent results can be obtained. All values of the second
derivative
$\sim\order{\grm/\mpl^2}$ preserve stability. The smallness of
$\vev{\tilde H_1}$ allows an easy solution to the condition for an
extremum in $S$ since the derivative of $\hat\mu$ appears only in
D-terms and Higgs mass terms which are proportional to $\tilde H_1^4$
and $\tilde H_1^2$ respectively, and hence make negligible
contributions to the determination of the minimum.  The fields
Re($S$) and Re($T$) have masses $\order{M_w^2/\mpl}$ while Im($S$)
and Im($T$) are massless at this level of analysis. These fields
could have important cosmological implications [11]. The remarkable
phenomenology given by this model is a consequence of string inspired
choices of boundary conditions and the tight constraints on the
location of the minimum.
\bigskip
\section{2. A Class of Globally Supersymmetric String-Inspired
Models}
Having illustrated a mechanism for determining a (global)
minimum in Higgs, dilaton and moduli fields, we broaden our
consideration to a class of globally supersymmetric models inspired by
the $4$-$D$ fermionic string construction. In so doing, we will be led to
a natural mechanism for the creation of the $\mu$ term, and will provide
motivation for the choice of rescaled $\mu$ and $\lamtop$ outlined in the
previous section.

The spectrum of fields can be assigned to string sets (twisted or
untwisted) and sectors (of which there are three in each set). The
K\"ahler potentials and soft SUSY breaking parameters
corresponding to the different choices are well enumerated
[12] and determine the necessary rescalings for the passage to the low
energy theory. Conservation of charge on the string world-sheet
determines the allowed couplings between different sets and sectors,
and hence restricts the possible soft terms which can arise [13]. In
particular, the analysis of [12] indicates that, whatever the
trilinear couplings present in the superpotential:
$$A=\grm,\quad\gam=\grm.\eqno(14)$$ The natural generalization of (6)
can be written
$$B=\Bigl(B_n - (S+\bar S){\partial\ln\mu\over\partial
S}\Bigr)\grm\quad{\rm with}\quad -n\leq B_n\leq n\eqno(15)$$
where $B_n$ arises from field differentials of the K\"ahler
potential (and may be $T$ dependent) and string-inspired potentials
restrict its possible value. Assignments of fields to sectors and
sets give, in general, non-universal values for
$m_0$, therefore although any particular scalar field will be massless or
of mass
$\grm$, we expect that choosing
$${m_0\over\grm}=0,1\ \ {\rm(universal)}\eqno(16)$$
will cover the
range of possible assignments as far as the results of the
minimization go. The choices $(14),\ (15)$ and $(16)$ define a class
of models suitable for the minimization procedure of Section 1, when
supplemented with suitable boundary conditions for $\lamtop$ and
$\mu$.

Field normalization and the rescaling of the superpotential
required to emulate a globally supersymmetric theory contribute to
an effective rescaling of all couplings present in the
superpotential [14]. With a $T$-independent string-derived Yukawa,
the restrictions on allowed couplings ensure that the
low-energy Yukawas are also $T$-independent. Further, treating the
origin of the $\mu$-term as a non-renormalizable term in the
superpotential of the form [15]
$$W_{nr}=\lambda_n\phi^{n-2}H_1H_2\quad\Rightarrow\quad
\mu=\lambda_n\vev{\phi}^{n-2}.\eqno(17)$$
It is clear that the rescaling of $\mu$ will depend on assignments of
the scalar fields $H_1$, $H_2$ and $\phi$. Restricting attention to
allowed couplings, we can often arrange for the rescaling not to
introduce any factors of $T$ into $\hat\mu$ or $B_n$, and hence we
can appeal to this freedom in justifying the choice made in Section
1. We choose $$\hat\mu=g\lambda_n\vev{\hat\phi}^{n-2}\eqno(18)$$ as
typical of the possible rescalings. The notation $\vev{\phi}$ is
purposefully vague since we consider the specification of $\phi$ to
be beyond the scope of this paper.

In Figures 1 and 2, we treat $B$ as an unconstrained parameter
coming from a string model and show the variation of the top mass and
lightest Higgs mass, of $\tan\beta$ and of the value of $\hat\mu$ required
to enforce an extremum. We note that since the $m_0=0$
and $m_0=\grm$ contours in Figures 1 and 2 are almost identical, we
expect that models with non-universal boundary conditions for $m_0$ will
give similar results. For all values of $B$, we obtain
$$\grm=1.5\ \tev,\quad 2.2\ \tev\quad{\rm for}\quad m_0=0,\quad\grm\quad{\rm
respectively}.\eqno(19)$$

At a deeper level of analysis, we treat $B_n$ as the defining parameter
coming from a string model, and illustrate in Figure 3 the range of $B$
for which $\partial\hat\mu/\partial S$ gives a stable theory. Again,
all reasonable values of the second derivative of $\hat\mu$ preserve
stability.

A further step comes from noting that with a string-derived coupling between
$n$ string fields with vertex correlation function $C$ [13,16]
$$\lambda_n=C{\sqrt{2}\over{(2\pi)^{n-3}}}g^{n-2}\eqno(20)$$
we can infer from $(18)$ that
$${\partial\hat\mu\over\partial S} =-\Bigl({{n-1}\over
2S}\Bigr){\hat\mu\over\mpl}\eqno(21)$$
where we have restored mass units. Using equation (15), Figure 3 shows
contours of constant $n$ which indicate the allowed form of the
non-renormalizable term in $W$.
\bigskip
\section{3. Conclusions and Speculations}
Having presented a simple scheme for the dynamical
determination of the vevs of all scalar fields in a realistic
no-scale model, we are in a position to speculate on the validity
and implications of our results. Having included only the top-Yukawa
and gauge couplings in the RGEs restricts the range of validity to
$\tan\beta<8$ [17]. However, since our results fall in this range we can
feel confident that the inclusion of other Yukawas will not
dramatically affect our results. We expect inclusion of the
whole spectrum of particles in the supertrace will quantitatively
change the results obtained. However this will not
invalidate the procedure since $V\rightarrow 0$ in the weak
coupling limit ($S\rightarrow\infty$), and $V\rightarrow\infty$
as $\alpha_3$ blows up ($S\rightarrow1.27$) -- a result which depends
on the squark and gluino contributions only. If $V<0$ for some $S>1.27$ as
in the examples presented, there will be a minimum.  A more serious question is
raised by the large value for $\grm$, for which a one-loop
corrected potential may not be sufficient [18]. However recent work [19] may
indicate that a scale well over $1\ \tev$ is permissible without
having to invoke a large degree of fine-tuning.

Comments in Section 1 regarding the implications of small values of
the (dimensionless) Higgs vevs also indicate that the inclusion of
moduli dependence in $\mu$ or $B$ will not change our procedure for
determining a minimum since $T$ dependence in $V$ beyond $\grm$ will
appear in terms multiplied by $\tilde H^2$. However, this
additional dependence will put an upper bound on the vev of $T$ since
stability of the theory may be affected by large values of
$\vev{T}$. Additional moduli dependence in the gauge couplings implied by
modular
invariance [20] appears at one-loop level and will again be negligible.

We note that in principle, string considerations will predict values for $\eta$
and $\mu$ which must reproduce the correct vev for $S$. However,
since string-derived values for these parameters are uncertain,
we choose to take the vev of $S$ as known and to regard the values
for the parameters required by our procedure as the values which
need to be derived from the string. In this way, we can regard these
values of $\eta$ and $\mu$ as tests of consistency rather than
predictions.

One may question the validity of trying to determine the vev of the
dilaton field using only the electro-weak scalar potential. Indeed,
realistic string models have extra $U(1)$ or GUT generators broken at
a large scale which will presumably dominate the minimization. In
principle, a similar analysis is possible including GUT structure.
However, the resulting vacuum energy will naturally be $\order{M^4_{\rm
GUT}}$ unless some mechanism is found to tame it.

The discrepancy between a string unification scale generically
$\order{10^{17}\ \gev}$ and a unification scale $\order{10^{16}\
\gev}$ extrapolated from low energy couplings in the SSM motivates
a string-inspired Standard Model with minimal extra particle content to
reconcile these scales [21]. The resulting modification of the RGEs for gauge
couplings will alter both the value of $g(\mx)$ and the dependence of the
various parameters in the low energy Lagrangian on $S$. This will probably
result in major quantitative differences in the results of our mechanism.

The critical assumption in this procedure has been to
break supersymmetry at tree-level with a constant term in the
superpotential. However, in the light of comments in Section 3
regarding the origin of a $\mu$-term, it is clear from equation
$(4)$ that once a $\mu$-term has been created through a singlet
field acquiring a vev, $W$ also acquires a vev dynamically. This provides a
mechanism for the dynamical breaking of supersymmetry without the tree-level
contribution. It is unclear how to reconcile such a supersymmetry
breaking mechanism with string no-go theorems. Moreover, a phenomenological
difficulty would be that the inferred vev of the modulus field will be
$\order{10^{-12}{\rm eV}}$ and hence we face a decompactification disaster.
However it is possible that an analysis of GUT structure and creation of a
GUT analogue of the $\mu$-term could overcome this problem. \bigskip
{\noindent\bf Acknowledgments}
\bigskip
One of us (S.\ K.) warmly thanks Dimitri Nanopoulos for support and
inspiration leading to the development of these ideas.

\vfil\eject

\def\refn #1;#2;#3;#4;#5;#6{\bigskip\item{[#1]}#2, #3 {\bf #4} (#5)
#6}
\def\ref #1;#2;#3;#4;#5{#1, #2 {\bf #3} (#4) #5}

\def\PLB{Phys.\ Lett.\ B}
\def\NPB{Nucl.\ Phys.\ B}
\def\DVN{D.\ V.\ Nanopoulos}
\def\JLL{J.\ L.\ Lopez}
\def\CK{C.\ Kounnas}

\section{References}

\leftskip40pt
\parindent 0pt

\refn 1;G.\ Veneziano, S.\ Yankielowicz;\PLB;113;1982;231;\ \ref T.R.\
Taylor;\PLB;165;1985;43; \ref S.\ Ferrara, N.\ Magnoli, T.\ R.\ Taylor, G.\
Veneziano;\PLB;245;1990;409; \ref B.\ de Carlos, J.A.\ Casas, C.\
Mu\~noz;\NPB;399;1993;623.

\refn 2;Particle Data Group;Phys.\ Rev.\ D;50;1994;.

\refn 3;R.\ Barbieri, G.\ Guidice;\NPB;306;1988;63.

\refn 4;E.\ Cremmer, S.\ Ferrara, \CK, \DVN;\PLB;133;1983;61.

\refn 5;E.\ Witten;\PLB;155;1985;51.

\refn 6;C.\ Kounnas, M.\ Porrati;\NPB;310;1988;355; \ref S.\ Ferrara, \CK, M.\
Porrati, F.\ Zwirner;\NPB;318;1989;75; \ref M.\ Porrati, F.\
Zwirner;\NPB;326;1989;162.

\refn 7;S.\ Kelley, \JLL, \DVN, H.\ Pois, K.\ Yuan;\NPB;398;1993;3.

\refn 8;M.\ Drees, M.\ M.\ Nojiri;\NPB;369;1992;54.

\refn 9; A.\ E.\ Faraggi, J.\ S.\ Hagelin, S.\ Kelley, \DVN;Phys.\ Rev.\
D;45;1992;3272.
\bigskip
\item{[10]}\DVN, ``The march towards no-scale supergravity'', CERN preprint
CERN-TH.7423/94 (hep-ph/9411281).

\refn 11;T.\ Banks, D.\ B.\ Kaplan, A.\ E.\ Nelson;Phys.\ Rev.\ D;49;1994;779;
B.\ de Carlos, J.\ A.\ Casas, F.\ Quevado, E.\ Roulet;\PLB;318;1993;447.
\bigskip
\item{[12]}\JLL, \DVN, CERN-TH.7519/94 (hep-ph/9412332).

\refn 13;S.\ Kalara, \JLL, \DVN;\NPB;353;1991;650.

\refn 14;S.\ K.\ Soni, H.\ K.\ Weldon;\PLB;126;1983;215.

\refn 15;J.\ E.\ Kim, H.\ P.\ Nilles;\PLB;138;1984;150; \ref G.\ F.\ Guidice,
A.\
Masiero;\PLB;206;1988;480.

\refn 16;S.\ Kalara, \JLL, \DVN;\PLB;245;1990;421.

\refn 17;M.\ Matsumoto, J.\ Arafune, H.\ Tanaka, K.\ Shiraishi;Phys.\ Rev.\
D;46;1992;3966.

\refn 18;E.\ Witten;\NPB;188;1981;513; \ref R.\ K.\ Kaul;\PLB;109;1982;19.

\refn 19; G.\ W.\ Anderson, D.\ J.\ Casta\~no;\PLB;347;1995;300

\refn 20; A.\ Font, L.\ E.\ Ib\'a\~nez, D.\ L\"ust, F.\
Quevado;\PLB;245;1990;401.

\refn 21;S.\ Kelley, \JLL, \DVN;\PLB;278;1992;140.

\leftskip 0pt

\vfil\eject

$$\def\vgap{height 3pt&&&&&&&&&&&&&&&&\cr}
\vcenter{\offinterlineskip\halign{
  &\vrule#&\kern.7em\hfil$\displaystyle{#}$\hfil\kern.7em\cr
  \noalign{\hrule}\vgap

&B&&\hat\mu&&\vev{H_1}&&\tan\beta&&\grm&&\mtop({\rm pole})
  &&m_{h}&&m_{H}&\cr
  \vgap\noalign{\hrule}\vgap
  &2.8\grm&&2.7\grm&&0.033\grm&&2.2&&2.2\ \tev && 165\ \gev
   &&60\ \gev&&7.0\ \tev&\cr
  \vgap\noalign{\hrule}\vgap
  &3.7\grm&&3.2\grm&&0.043\grm&&1.6&&2.2\ \tev&&153\ \gev
   &&38\ \gev&&8.7\ \tev&\cr
  \vgap\noalign{\hrule}
}}$$
\centerline{Table 1. Range of consistent results for minimization of the
model $m_0=1$, $A=3$.}

\vfil\eject

\section{Figure Captions}
\leftskip40pt
\parindent 0pt

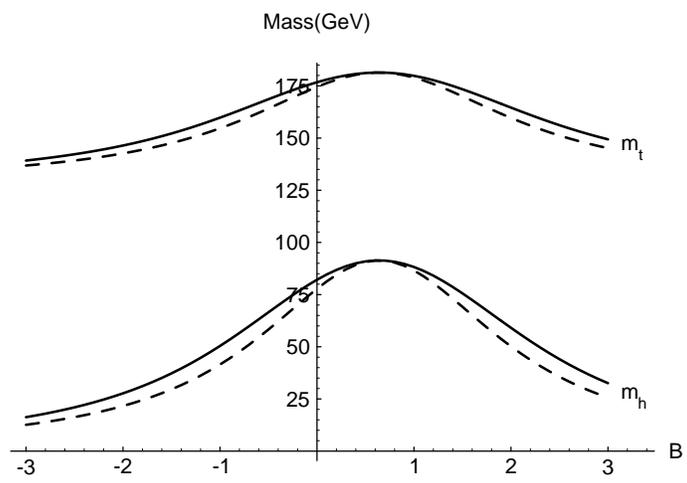
\item{Fig. 1.}
Masses of the Top quark and lightest Higgs for the class of model in Section
2 with $B$ (in units of $\grm$) as a parameter. Dashed lines indicate
$m_0=0$, solid lines indicate the case $m_0=\grm$. \bigskip
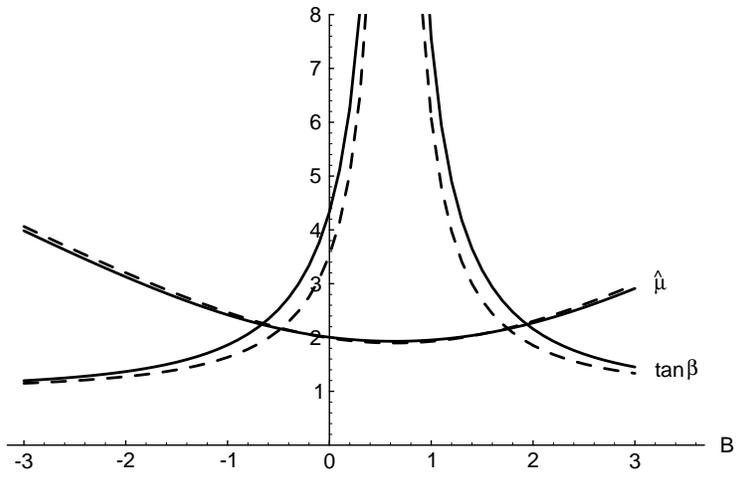
\item{Fig .2.}
Values of $\tan\beta$ and $\hat\mu$ (in units of $\grm$) at the minimum
for the class of model in Section 2 with $B$ (in units of $\grm$) as a
parameter. Dashed lines indicate $m_0=0$, solid lines indicate the case
$m_0=\grm$. \bigskip
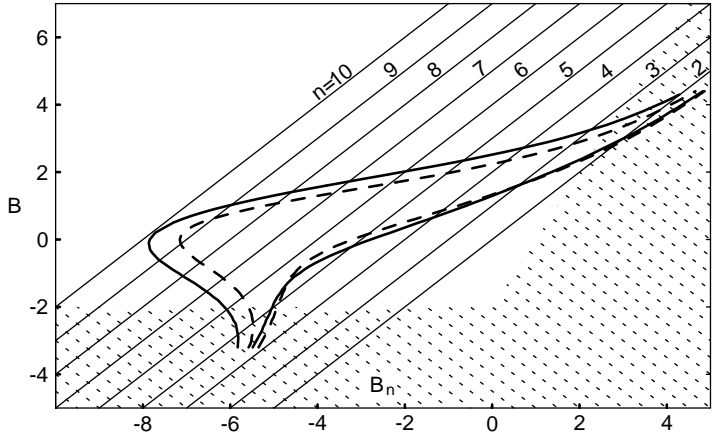
\item{Fig. 3.}
Range of stability of a model defined by a value of $B_n$. Solid
diagonal lines indicate the number of fields $n$ in a non-renormalizable
term which create the required dilaton dependence in a $\mu$-term. The
region between dashed lines has $m_0=0$ and between solid lines has
$m_0=\grm$. The dotted region is excluded by equation $(15)$. $B$ is
measured in units of $\grm$.

\vfil\eject
\leftskip 0pt

$$\bpict massplot (3.92in by 2.42in)
labels \put r {Mass ($\gev$)} at (1.6in,2.5in)
\put r $B$ at (3.95in,0.09in)
\put r $\mtop$ at (3.9in,1.9in)
\put r $m_h$ at (3.9in,0.5in)
\put r {Figure 1.}  at (1.7in,-0.4in)
left 0pt right 0pt up 0pt
down 0pt sc 1000 epict $$

$$\bpict tanbplot (3.92in by 2.42in)
labels \put r $B$ at (3.85in,0.07in)
\put r $\tan\beta$ at (3.8in,0.47in)
\put r $\hat\mu$ at (3.8in,0.92in)
\put r {Figure 2.}  at (1.7in,-0.4in)
left 0pt right 0pt up 50pt down 0pt sc
1000 epict $$

\eject

$$\bpict BnBplot (3.92in by 2.42in)
labels \put r $B$ at (-0.17in,1.25in)
\put r $B_n$ at (2.0in,-0.2in)
\put r {Figure 3.}  at (1.7in,-0.6in)
left 0pt right 0pt up 0pt down 0pt sc
1000 epict $$

\bye